# Thinking Twice: A Sequential Approach to Nudge Towards Reflective Judgment in GenAI-Assisted Decision Making

**Research Paper**

Hüseyin Hussein Keke[1], Daniel Eisenhardt[1], Christian Meske[1]

[1] Ruhr University Bochum, Chair of Socio-technical System Design & AI, Bochum, Germany
{hueseyin.husseinkeke, daniel.eisenhardt, christian.meske}@ruhr-uni-bochum.de

**Abstract.** Humans increasingly use generative AI (GenAI) for decision-making assistance, yet tend to engage in such situations using quick, intuitive (System 1) rather than deliberate, analytical assessment (System 2) as reflected in Dual Process Theory. Such engagement reduces the effectiveness of human-GenAI interactions and increases associated risks. Our research investigates how to design human-GenAI interactions that promote the transition from System 1 to System 2 thinking in decision-making. Based on the Three-Stage-Model, which suggests humans switch to System 2 when conflicts are detected, we conducted an experiment (n=130) testing sequential human-GenAI interaction as a cognitive forcing nudge during problem-solving tasks. This approach—where humans decide first before receiving AI assistance—enables detection of conflicting decisions. Compared to non-sequential settings (concurrent or no AI assistance), results show the sequential condition significantly enhanced both decision performance and GenAI prompt utilization, providing valuable insights for designing more reflective and thus effective human-GenAI interactions.

Please note that the review process is double-blind. Manuscripts submitted for review MUST NOT include author information—neither on the title page nor in the page header, etc. This paragraph is intended to serve as filler in your initial submission to keep the overall length of the paper consistent, even if the author information section above is not included. You can delete it in the final submission when you add the author information.

## 1 Introduction

The interplay between human cognition and artificial intelligence (AI) presents both unprecedented opportunities and complex challenges in decision-making processes (Steyvers & Kumar, 2024; Wei et al., 2025). As generative AI (GenAI) systems become increasingly sophisticated and accessible, understanding how humans cognitively engage with these tools becomes essential for maximizing their benefit. Research at this intersection reveals a critical gap: our tendency to process AI-generated information through quick, intuitive judgments rather than deliberate analytical reasoning (Hao et al., 2024). Humans' natural reliance on intuitive cognition often leads to systematic biases in decision-making processes (Kahneman, 2011). While such heuristic approaches offer efficiency through rapid, low-effort processing and generally yield satisfactory outcomes, they can result in suboptimal decisions when logical considerations are warranted. This cognitive architecture is explained by Dual Process Theory (DPT), which delineates two distinct modes of thought: System 1, characterized by fast, automatic processing, and System 2, marked by deliberate, analytical reasoning (Kahneman, 2011). The engagement of System 2 becomes particularly crucial in decision contexts with significant implications, where reliance on intuitive processing alone may prove insufficient for optimal outcomes (Janssen et al., 2020).

This tension between intuitive and analytical thinking takes on new dimensions as technological tools evolve to support human decision-making. In this context, the rapid advancement of GenAI represents a fundamental shift in decision support capabilities. While GenAI tools like ChatGPT demonstrate significant potential to improve decision-making across multiple fields (Feuerriegel et al., 2024; Huy et al., 2024), a critical challenge remains: even when these systems provide accurate guidance, users often default to their intuitive thinking rather than engaging in critical analysis of AI-generated insights (Litvinova et al., 2024). This tendency to rely on System 1 thinking, even when AI support is available, may prevent the full realization of GenAI's potential to enhance decision performance. Research suggests that the presentation of AI-generated outputs plays a crucial role in shaping user engagement and promoting more analytical thinking (Carter & Liu, 2025; Rosenbacke et al., 2024).

Digital nudges and cognitive forcing functions (CFFs) represent complementary design approaches to mitigate this challenge. Nudges guide user behavior without restricting choice, while CFFs aim to interrupt automatic processing and trigger deliberate reasoning. While these approaches have shown promise (e.g., Green & Chen, 2019; Zhu et al., 2025), current IS research has primarily focused on their application to explainable AI (XAI) systems (e.g., Bertrand et al., 2022), with limited investigation of how they might enhance analytical engagement with GenAI (Hao et al., 2024). Despite recent mentions in the literature (e.g., Rosenbacke et al., 2024), an effective way to implement these approaches in GenAI interfaces has yet to be developed. To address this gap, we turn to the Three-Stage Model (TSM) (Pennycook et al., 2015), which provides a theoretical lens for understanding how users can be nudged from System 1 to System 2 thinking by detecting conflicts between intuitive responses and subsequent information. GenAI interaction designs may create opportunities for conflict detection, enhancing analytical engagement. Thus, we propose:

**RQ**: How can human-GenAI interactions be designed to facilitate the transition from System 1 to System 2 thinking in decision-making tasks?

To address this RQ, we tested sequential human-GenAI interaction as a cognitive forcing nudge via a controlled experiment (n=130) involving problem-solving tasks (brain teasers). Amongst others, our findings reveal that having users decide first before receiving AI assistance increased critical reflection and significantly improved decision performance as well as increased prompt utilization compared to concurrent and hence non-sequential AI assistance, or no AI assistance. This paper is structured as follows: we review existing research on GenAI-assisted decision-making (Section 2) and establish our theoretical framework and hypotheses (Section 3). We then describe our methodology (Section 4). Section 5 presents our analysis, followed by a discussion (Section 6) and conclusion, including limitations and outlook to further research (Section 7).

## 2 Related Work

AI-assisted decision-making creates a complex dynamic where individuals must effectively balance and incorporate AI outputs alongside their own reasoning and evaluation processes. In such scenarios, AI provides predictions and/or explanations to assist the human decision-maker, who remains responsible for the final choice (Rosenbacke et al., 2024). Recent research on human-AI decision-making has therefore focused on understanding and optimizing this collaborative performance (Buçinca et al., 2020; Schemmer et al., 2023) across various domains, including clinical diagnosis, financial decisions, and forecasting (Steyvers & Kumar, 2024). This human-AI collaboration requires carefully designed workflows and systems that align with human needs, enabling effective collaboration with AI (Carter & Liu, 2025; Huy et al., 2024).

Recent IS research focused on AI-assisted decision-making, emphasizing machine learning-based discriminative AI for classification and prediction, including predictive models, recommender systems, and sentiment analysis (e.g., Stanciu et al., 2025; Zhou et al., 2024; Zhou et al., 2022). Scholars enhance these Decision Support Systems (DSS) with a primary focus on XAI explanations to help users make more informed decisions by revealing underlying decision processes (Ehsan et al., 2024; Meske et al., 2022). Understanding how AI explanations interact with human cognitive processes requires careful consideration of various cognitive biases (e.g., mere exposure effect, anchoring bias, faith in numbers) (Bertrand et al., 2022; Carter & Liu, 2025; Ehsan et al., 2024). DPT provides a framework for understanding these interactions, and while research in this area remains limited (Litvinova et al., 2024; Lu & Zhang, 2024), interventions like CFFs build on DPT principles to promote analytical engagement with AI outputs, helping to overcome cognitive bias and fostering more effective human-AI interaction (Green & Chen, 2019; Rosenbacke et al., 2024).

Unlike AI systems designed for discriminative tasks, GenAI systems, especially Large Language Models (LLMs) like ChatGPT, represent a new paradigm requiring further study. Their ability to generate content and engage in natural dialogue creates novel challenges for effective collaboration (Feuerriegel et al., 2024; Huy et al., 2024). GenAI has been explored in a wide range of topics, including IT recruitment (Szandała, 2025), scholarly communication (Davazdahemami et al., 2024), word-of-mouth in hospitality (Mladenović et al., 2024), corporate performance (Ye et al., 2024), product discovery (Gude, 2024), technical bias (Wei et al., 2025) and creative problem-solving (Dwivedi & Banerjee, 2024). However, current IS research has limited exploration of cognitive aspects in human-GenAI collaboration (e.g., Sun & Kok, 2024), particularly regarding heuristics and cognitive biases. Notably, when familiar cognitive patterns are absent, users tend to default to intuitive rather than analytical processing in GenAI interactions (Hao et al., 2024).

To deepen the understanding of cognitive processing during human-GenAI collaboration, a promising theoretical approach integrates the DPT and the TSM capturing the cognitive switch between System 1 and System 2 (Pennycook et al., 2015). This approach, beyond its previous applications in XAI, can also be applied to LLM contexts (Litvinova et al., 2024), potentially enhancing our understanding of cognitive processing. However, explainability research of LLM models remains in nascent stages, and potential interventions or design modifications specific to LLM contexts have yet to be examined. The following sections will address this research gap in detail.

## 3 Theoretical Background and Hypothesis Development

DPT explains human reasoning through two cognitive systems: System 1's rapid, automatic processing and System 2's deliberate, analytical thinking (Kahneman, 2011). Building on this, the TSM by Pennycook et al. (2015) details how these systems interact during decision-making. In Stage 1, System 1 processes generate multiple intuitive responses with varying emergence speeds, e.g., stereotype-based responses typically emerge faster than statistical reasoning. Stage 2 involves conflict monitoring, screening for discrepancies between these initial responses. Without detected conflicts (either due to absence or monitoring failure), the dominant intuitive response proceeds to Stage 3 with minimal analytical evaluation, often leading to biased judgments. When conflicts are detected, more substantial System 2 processing is triggered in one of two forms: rationalization or cognitive decoupling. Rationalization uses System 2 processing to justify the initial intuitive response without considering alternatives. In contrast, cognitive decoupling involves actively inhibiting the initial response to consider alternatives or generate new solutions. This model integrates metacognitive monitoring, conflict detection, and response inhibition mechanisms, providing a framework for understanding both analytical thinking triggers and individual variations in reasoning processes.

Building on the DPT, scholars have utilized nudging to design decision architectures that subtly guide behavior (Thaler & Sunstein, 2003). In the context of human-centered AI, digital nudging has evolved across discriminative AI applications (e.g., Wang et al., 2022), progressing from “traditional” nudges (e.g., Pálfi et al., 2024; Wang et al., 2022) to AI-driven interventions (e.g., Duane et al., 2024; Sadeghian et al., 2023; Sumner et al., 2023). These applications span various domains from healthcare (e.g., Sumner et al., 2023; Pálfi et al., 2024) to education (e.g., Zhu et al., 2025), though decision-making with GenAI remains understudied (Hao et al., 2024). As AI systems increasingly influence decision-making (Zhu et al., 2025), a fundamental aspect of nudging to consider is that interventions must be transparent and preserve freedom of choice, especially when designed to promote analytical thinking (Meske & Amojo, 2020). Particularly promising could be structural nudges that alter decision processes through increased effort, with Jesse and Jannach's (2021) classification highlighting mechanisms like friction (Caraban et al., 2019) and "speed bumps" (Sunstein, 2016). These interventions aim to encourage mindful behavior and activate reflective System 2 thinking (Caraban et al., 2019; Leimstädtner et al., 2023; Sundin, 2021). A concept aimed at influencing thinking processes in decision-making situations, which also parallels these mechanisms, are CFFs (Buçinca et al., 2021). The common ground between the concepts is the understanding DPT and the possibility to nudge individuals from System 1 to System 2 to make thoughtful decisions, counteracting heuristic reasoning (Lambe et al., 2016). The distinction is that no claim to decision-making freedom is made, as "forcing" already implies a lack of choice. Examples include requiring users to make a decision before receiving AI assistance, introducing a time delay before AI support is provided, or providing assistance only upon request (Bertrand et al., 2022; Park et al., 2019; Steyvers & Kumar, 2024).

Based on this, we propose hypotheses and examine the influence of nudging on performance in GenAI-assisted cognitively demanding tasks, specifically brain teaser puzzles that require analytical thinking. Individuals often rely on System 1 processing, leading to quick but potentially incorrect intuitive responses (Kahneman, 2011). The introduction of AI assistance can serve as an external possibility for conflict detection, helping users identify discrepancies between their intuitive responses and alternative solutions (Abdel-Karim et al., 2023; Hao et al., 2024). By providing concurrent AI assistance available on demand, it may facilitate the transition to System 2 processing, enabling cognitive decoupling from initial intuitive responses or rationalization of the initial intuitive response (Bertrand et al., 2022; Pennycook et al., 2015). Also, recent research has shown that user performance can improve when users have the opportunity to use (Gen)AI assistance, in comparison to performance without (Gen)AI (Lu & Zhang, 2024; Mehler & Krautter, 2024). Hence, we hypothesize:

**H1:** Users receiving concurrent GenAI assistance will demonstrate higher performance on brain teaser tasks compared to users without such assistance.

Moreover, the literature on AI-assisted decision-making examines the optimal timing for providing support to optimize human-AI collaboration. Scholars have investigated how different timing points for AI assistance can nudge users to engage with the AI output, with the sequential approach demonstrating potential (e.g., Rosenbacke et al., 2024). In this approach, the AI's output is shown after the user's initial decision, with the option to modify the decision based on the AI's input (Lambe et al., 2016). Since users in the sequential condition have access to the same GenAI support as the concurrent condition, differing only an initial unassisted decision, similar cognitive decoupling or rationalization mechanisms could occur (Buçinca et al., 2021; Pennycook et al., 2015). Research also suggests that the sequential approach enhances the accuracy of AI-assisted decisions in comparison to human performance (Abdel-Karim et al., 2023), likely by encouraging independent reflection and aiding in the retrieval of relevant information (Green & Chen, 2019; Steyvers & Kumar, 2024). Therefore, we hypothesize:

**H2:** Users receiving sequential GenAI assistance will demonstrate higher performance on brain teaser tasks compared to users without such assistance.

In addition to these mechanisms, from the TSM perspective, the sequential approach may align with theoretical assumptions by providing a structured reflection opportunity before AI assistance. This sequential arrangement potentially enables the activation of endogenous conflict detection mechanisms (Abdel-Karim et al., 2023; Pennycook et al., 2015) while also utilizing AI as an external mechanism for conflict detection (Abdel-Karim et al., 2023; Hao et al., 2024; Litvinova et al., 2024). The temporal and structural separation between initial processing and AI assistance might enhance the effectiveness of these processes (Litvinova et al., 2024; Park et al., 2019), thereby reflecting mentioned nudges such "speed bumps" (Sunstein, 2016) or friction nudges (Caraban et al., 2019) as well as CFFs (e.g., Buçinca et al., 2021). Thus, we hypothesize:

**H3:** Users receiving sequential GenAI assistance will demonstrate higher performance on brain teaser tasks compared to users receiving concurrent GenAI assistance.

Building on the sequential approach, we assume that this approach not only improves user performance compared to concurrent AI assistance, but also that performance is enhanced *within the group* using the sequential approach. On one hand, scholars have shown that AI assistance leads to better performance (Lu & Zhang, 2024; Mehler & Krautter, 2024), and on the other hand the mentioned cognitive mechanisms based on TSM should occur during the decision process as users transition from initial intuitive responses to analytical ones through this structured approach, activating conflict detection and leveraging external AI resources (Abdel-Karim et al., 2023; Green & Chen, 2019; Hao et al., 2024; Pennycook et al., 2015). Consequently, we hypothesize:

**H4:** Users receiving sequential GenAI assistance will demonstrate performance improvement from their initial unassisted responses to their subsequent GenAI-assisted responses on brain teaser tasks.

Building on the assumption that the sequential approach may facilitate the retrieval of relevant information (Abdel-Karim et al., 2023; Steyvers & Kumar, 2024), which in this context refers to the use of external AI resources, while also offering an additional mechanism for conflict monitoring (Abdel-Karim et al., 2023; Litvinova et al., 2024; Pennycook et al., 2015), this may, in turn, encourage higher prompt utilization as users actively work to resolve cognitive conflicts identified during their initial attempt. Thus, we hypothesize:

**H5:** Users receiving sequential GenAI assistance will demonstrate significantly higher prompt utilization compared to users receiving concurrent GenAI assistance.

# 4 Method

## 4.1 Research Design

An online experiment was designed to assess how a cognitive forcing nudge affects participants' engagement with cognitively demanding tasks using ChatGPT. Participants were randomly assigned to three conditions: a control group without assistance and two experimental groups with different ChatGPT interaction designs. The task involved solving brain teasers, puzzles that typically trigger intuitive but incorrect first responses, requiring analytical thinking for their solution. Two well-known problems were selected: the bat-and-ball problem (Janssen et al., 2020) and the Monty Hall problem (Wilcox, 2024), chosen for their tendency to elicit initial misunderstandings, emphasizing the importance of analytical reasoning (Figure 1).

For each brain teaser, participants selected from five possible answers, with only one being correct. In the experimental conditions, participants had access to three predefined ChatGPT prompts designed to support correct answer selection. The interaction was conversational, with ChatGPT's accurate responses appearing after a brief loading animation when participants double-clicked a prompt.

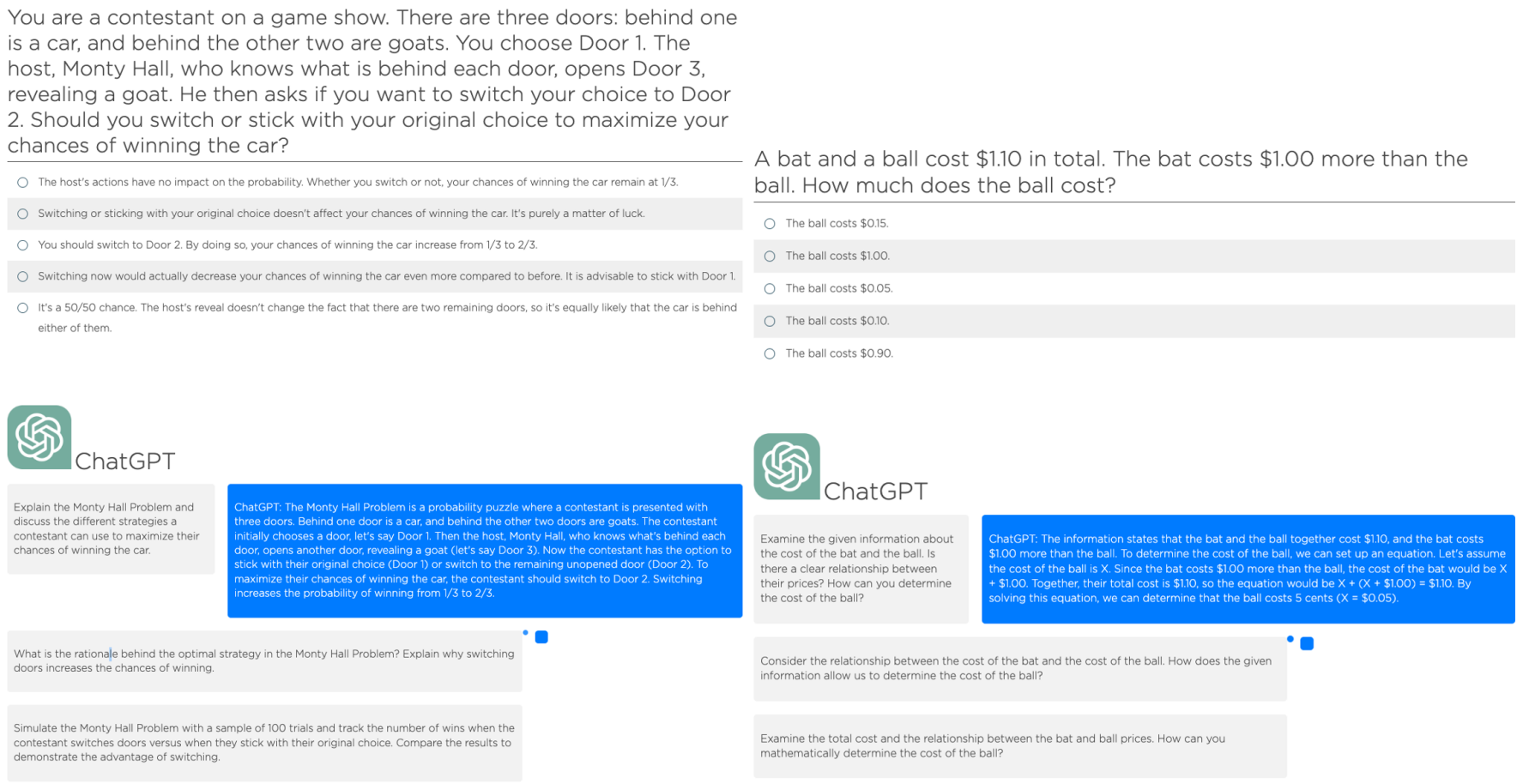

**Figure 1.** *Brain Teasers (Left: Monty Hall Problem; Right: Bat-and-ball problem)*

In the concurrent ChatGPT assistance experimental group, participants clicked on the prompt and immediately received a response from ChatGPT. In the sequential ChatGPT assistance experimental group, we implemented a sequential process: participants first had the option to make a decision on their own. If they preferred, they could skip this initial step and proceed directly to the same brain teaser with ChatGPT's assistance. The first decision was unassisted and optional, while the second decision was assisted and mandatory. This design draws on CFFs, suggesting that a sequential approach can improve performance in AI-assisted decision-making (Green & Chen, 2019). To meet ethical nudging standards, the first decision remained optional, aligning with nudge theory's principle that nudges should be easy to avoid (Kahneman, 2011). Figure 2 illustrates the procedures within the groups and the corresponding hypotheses.

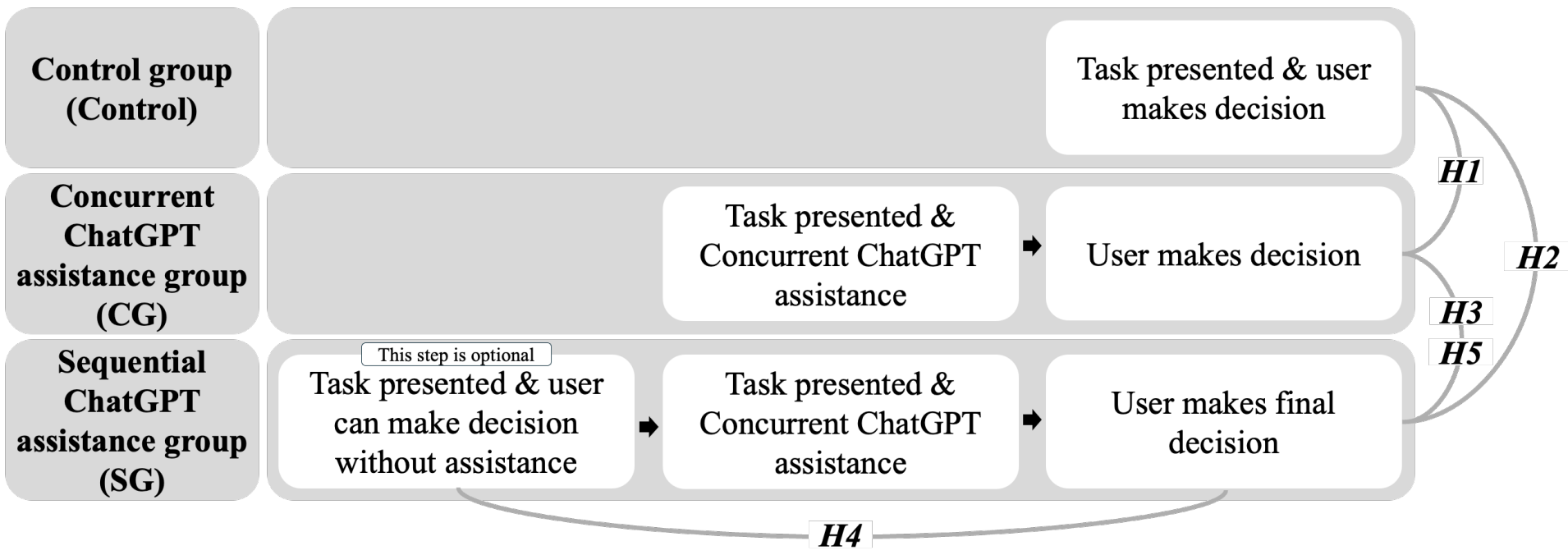

**Figure 2.** *Decision Structure & Hypothesis*

### 4.2 Sample and Statistical Analysis

The sample included 150 independently recruited participants from the Prolific platform. In the analysis of the sample, three participants were excluded for failing the attention check, and 17 participants were removed due to inconsistencies in their responses, such as reporting the use of ChatGPT without prompts usage or vice versa. This resulted in a final sample of 130 participants. All participants, proficient English speakers over 18 years old, born and residing in the UK, received financial compensation. Participants ranged in age from 19 to 75 years (*M* = 36.34 years, *SD* = 12.06), and 64.6 % of participants were female.

This study employed an ANOVA to examine treatment effects on performance, complemented by paired and independent t-tests to analyze sequential group performance differences and prompt usage between concurrent and sequential conditions.

## 5 Results

### 5.1 Descriptive Statistics

The initial section of the results provides an overview of the descriptive statistics. The means represent the average performance within the group, expressed as a percentage. The performance increased from control group (*N* = 50, *M* = 33.00, *SD* =37.27), to concurrent ChatGPT assistance group (*N* = 41, *M* = 48.78, *SD* =36.21), to sequential ChatGPT assistance group (*N*= 39, *M* = 66.67, *SD* = 35.04). Within the sequential ChatGPT assistance group, performance also increased, as the baseline at the first response was notably lower (*N* = 39, *M* = 38.46, *SD* = 33.37). The number of prompts used increased from the concurrent ChatGPT assistance group (26 prompts, *M* = 0.63, *SD* = 1.04) to the sequential ChatGPT assistance group (52 prompts, *M* = 1.33, *SD* = 1.34). The mean values and the prompts count are presented in figure 3.

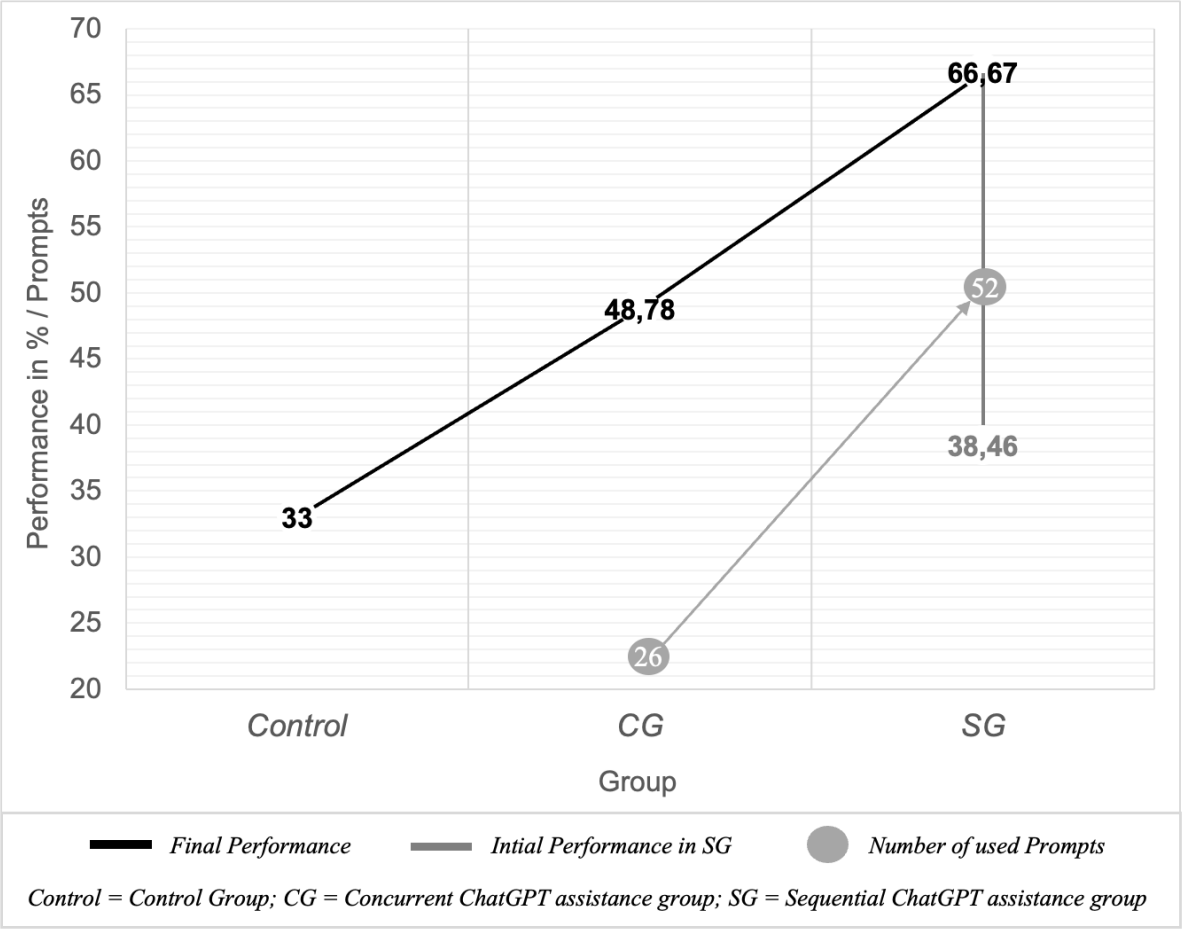

**Figure 3.** *Performance and Number of Prompts per Group*

## 5.2 ANOVA Analysis and Post-Hoc Analysis

The first step of the analysis was to assess the required assumptions for ANOVA analysis. There were no outliers, as indicated by a box plot check. Data was not normally distributed for each group (Shapiro-Wilk test, $p > .05$), which is not a major issue as an ANOVA is robust to violations of this criterion (e.g., Blanca et al., 2017). Homogeneity of variances was confirmed by applying the Levene test ($p = 0.245$). The results are sufficient to continue with the analysis.

The results of the ANOVA (Table 1) demonstrate a statistically significant difference in performance for the different groups, F (2, 127) = 9.46, $p$ = <.001. The effect size was between small to medium ($\eta^2 = 0.129$) (Cohen, 1988). The next step involves identifying the specific differences between the groups using post-hoc tests. The LSD post-hoc analysis (Table 1) revealed a significant difference in performance. Performance increased from the concurrent ChatGPT assistance group to the control group (15.78, 95% CI [-00.66, 30.91], $p = .041$). Performance also increased significantly from the sequential ChatGPT assistance group to the control group (33.67, 95% CI [18.33, 49.00], $p$ = <.001), and from the sequential ChatGPT assistance group to the concurrent ChatGPT assistance group (17.87, 95% CI [1.89, 33.94], $p = .029$). Thus, hypotheses H1, H2, and H3 can be confirmed.

**Table 1.** *ANOVA & Post-Hoc Analysis of Performance*

| | *ANOVA* | | | | *Pairwise Post-Hoc Analysis* | | | | |
|---|---|---|---|---|---|---|---|---|---|
| | **df** | **F** | **Sig.** | **$\eta^2$** | | **$M_{diff}$** | **SE** | **Sig.** | **95% CI** |
| Between Groups | 2 | 9.445 | *<.001**** | .129 | Control - CG | 15.78 | 7.64 | *.041** | [-00.66, 30.91] |
| Within Groups | 127 | | | | Control - SG | 33.67 | 7.75 | *<.001**** | [18.33, 49.00] |
| | | | | | CG - SG | 17.89 | 8.11 | *.029** | [1.89, 33.94] |

Control = Control group; CG = Concurrent ChatGPT assistance group; SG = Sequential ChatGPT assistance group; *p < .05, **p < .01, ***p < .001

## 5.3 Paired & Independent T-Test

The assumptions for the paired and independent T-test analysis (Table 2) were checked, with no (extreme) outliers identified via a box plot and non-normality assessed using the Shapiro-Wilk test ($p > .05$). Given the T-test's robustness to violations of normality (Stone, 2010a; Stone 2010b), the analysis can proceed. Homogeneity of variances for the independent t-test analysis was confirmed using the Levene test ($p = 0.305$).

A paired T-test was employed to assess the changes in performance within the sequential ChatGPT assistance group. The first response was optional, but except for one response in one task, all participants provided two responses. The analysis demonstrates that the performance within the sequential ChatGPT assistance group significantly improved from the first step to the second, t (38) = 5.18, $p$ = <.001. The effect size is large d = 0.83 (Cohen, 1988). H4 is supported.

An independent t-test was conducted to evaluate differences in prompt usage between the concurrent ChatGPT assistance group and the sequential ChatGPT assistance group. The analysis demonstrates that prompt usage was significantly higher in the sequential ChatGPT assistance group, t (78) = 2.61, $p$ = .011. The effect size is medium d = 0.58 (Cohen, 1988). H5 is supported. All results are presented in Table 3.

**Table 2.** *T-Tests*

| T-Test | Comparison | t | df | Sig. | 95% CI |
|---|---|---|---|---|---|
| Paired | Performance (SG 2 – SG 1) | 5.18 | 38 | *<.001*** | [17.18, 39.23] |
| Independent | Prompt Usage (CG – SG) | 2.61 | 78 | *.011** | [0.17, -1.23] |

CG = Concurrent ChatGPT assistance group; SG = Sequential ChatGPT assistance group (1= Initial & 2 = Final); *p < .05, **p < .01, ***p < .001

**Table 3.** *Overall Results*

| Hypothesis | DV | Support |
|---|---|---|
| **H1**: Control < CG | Performance | Yes |
| **H2**: Control < SG | Performance | Yes |
| **H3**: CG < SG | Performance | Yes |
| **H4**: SG 1 < SG 2 | Performance | Yes |
| **H5**: CG < SG | Prompts | Yes |

Control = Control group; CG = Concurrent ChatGPT assistance group; Sequential ChatGPT assistance group (1= Initial & 2 = Final); DV = Dependent Variable

# 6 Discussion

## 6.1 Nudging GenAI-Assisted Decision-Making

The experimental findings from our study offer insights into how different implementations of GenAI assistance affect human decision-making processes. By leveraging theoretical frameworks from Dual Process Theory (DPT) and the Three-Stage Model (TSM), we designed an investigation that combined digital nudging with cognitive forcing functions to examine sequential ChatGPT assistance during brain teaser tasks. Our analysis of 130 participants revealed several interesting patterns: first, the sequential AI interaction approach led to enhanced performance outcomes compared to both control and concurrent assistance conditions (see Figure 3); second, when participants were required to make initial independent attempts, they subsequently engaged more thoroughly with the AI system through increased prompt utilization; and third, we observed improvement in decision performance when comparing participants' initial responses to their final answers after AI engagement.

Participants in the sequential group significantly outperformed those in the control group without assistance (H2) and to concurrent assistance group (H3). Specifically, requiring participants to make an initial unassisted attempt appears to promote the transition from System 1 to System 2 thinking through enhanced conflict detection, confirming prior research (Abdel-Karim et al., 2023; Green & Chen, 2019; Litvinova et al., 2024; Pennycook et al., 2015). This may be explained by two key observations: First, participants showed significant improvement between their initial intuitive responses and subsequent analytical answers in the sequential condition (H4), indicating a successful shift from System 1 to System 2 processing. Second, those in the sequential condition demonstrated higher levels of prompt utilization compared to participants receiving concurrent assistance (H5), suggesting more deliberate engagement with the available information. The sequential design appears to help users identify discrepancies between their intuitive responses and alternative solutions through a temporal and structural separation (Green & Chen, 2019; Park et al., 2019; Pennycook et al., 2015; Steyvers & Kumar, 2024) that may make conflicts more salient and harder to ignore. This increased awareness likely motivates users to engage more thoroughly with GenAI assistance when it becomes available, while also providing an external opportunity for conflict detection (Abdel-Karim et al., 2023; Hao et al., 2024; Litvinova et al., 2024), thereby leveraging capabilities of computational power (Hao et al., 2024).

Overall, these findings validate our theoretical foundation utilizing DPT and TSM, and the integration of CFFs with nudging principles that preserve user autonomy. The demonstrated effectiveness of sequential AI assistance in enhancing decision accuracy and prompt utilization offers valuable theoretical and practical insights for the design of human-AI collaboration, which will be further elaborated in the following section.

## 6.2 Implications for Research and Practice

Our research enhances the understanding of AI-assisted decision-making by advancing theoretical perspectives and building upon existing literature. First, this research advances understanding by applying DPT and TSM to the domain of GenAI systems. While psychology has explored the shift between intuitive and analytical thinking (Kahneman, 2011; Pennycook et al., 2015), its application to AI-assisted decision-making, particularly in the context of GenAI, remains understudied (Hao et al., 2024; Litvinova et al., 2024). Our study addresses this theoretical gap by examining AI assistance's influence on cognitive processes through the TSM. Extending prior research (Hao et al., 2024; Litvinova et al., 2024), we provide an approach for understanding cognitive mechanisms in GenAI-assisted decision-making, differentiating the progression from intuitive responses through conflict detection to analytical judgments (Abdel-Karim et al., 2023; Hao et al., 2024; Pennycook, 2015). Second, we extend the digital nudging literature in IS by providing insights into decision structure-related and reflection-promoting nudges (Caraban et al., 2019; Sundin, 2021; Sunstein, 2016). While digital nudging has been studied in IS (Jesse & Jannach, 2021), theoretical approaches examining cognitive mechanisms remain limited in GenAI contexts (Hao et al., 2024). Third, we demonstrate the intersection between CFFs and digital nudging in GenAI interventions, particularly regarding their theoretical foundations (Kahneman, 2011; Lambe et al., 2016) and reflection-promoting approaches (Green & Chen, 2019; Leimstädtner et al., 2023). Our findings show that this synthesis preserves the effectiveness of CFFs (Rosenbacke et al., 2024) while maintaining nudging's emphasis on decision autonomy (Thaler & Sunstein, 2003).

Our research also provides practical insights for implementation. Organizations could consider designing their GenAI interfaces to incorporate structured decision steps rather than providing immediate AI assistance, particularly in high-stakes domains (Carter & Liu, 2025). Moreover, this can be achieved by implementing optional reflection phases before presenting AI suggestions, allowing users to document their initial thoughts (Green & Chen, 2019; Steyvers & Kumar, 2024). Moreover, such an approach could promote better performance while maintaining user autonomy (Thaler & Sunstein, 2003) through an opt-out mechanism, where users have the ability to skip the initial reflection phase rather than having AI assistance appear automatically. Additionally, this sequential approach may prove especially valuable in decision contexts where intuitive responses could potentially lead to incorrect conclusions (Hao et al., 2024; Janssen et al., 2020), possibly creating situations where users might dismiss the benefits of AI assistance. Consequently, the structured decision steps could help employees recognize potential gaps between their initial intuitive responses and more analytical solutions (Abdel-Karim et al., 2023; Green & Chen, 2019; Pennycook et al., 2015), extending, for example, Microsoft's considerations for GenAI interaction design (Passi et al., 2024). Therefore, our approach demonstrates that employees can be trained to utilize GenAI, enhancing their ability to make more informed decisions and optimize work outcomes through comprehensive collaboration monitoring and strategic implementation (Hao et al., 2024).

# 7 Conclusion, Limitations and Outlook to Further Research

Our study examines how different implementations of GenAI assistance influence decision-making quality by requiring users to make initial unassisted attempts before receiving AI support. Drawing on Dual-Process Theory (DPT) and the Three-Stage Model (TSM), we combine sequential nudging interventions with cognitive forcing functions (CFFs) in our experimental design. Our findings demonstrate that sequential AI assistance improves performance on brain teaser tasks compared to concurrent assistance, with users showing enhanced prompt engagement and better decision outcomes when required to make initial unassisted decisions. These results validate the effectiveness of integrating nudging principles with cognitive forcing functions while maintaining user autonomy in AI-assisted decision processes.

As with every study, this research has limitations that suggest directions for future research. First, our study used pre-constructed ChatGPT prompts instead of live API integration and brain teasers rather than real-world decisions. This ensured cost-effectiveness and controlled conditions but may have limited the authenticity of user-AI interaction. Future research could integrate the ChatGPT API and real-world tasks to explore dynamic AI interactions' impact on decision-making and user engagement. Second, we focused on performance and prompt differences as observable indicators of the nudge's impact, theoretically grounding our approach in DPT and TSM without direct assessments. Future studies could enhance this foundation by incorporating measures to strengthen the theoretical foundation (e.g., Pennycook et al., 2015; Song et al., 2021). Third, while our focus on intervention design with AI providing helpful answers offered valuable insights, it necessarily excluded examination of related phenomena such as AI error handling, overreliance, and automation bias (e.g., Buçinca et al. 2021; Litvinova et al. 2024). Future research could extend our experimental paradigm to investigate these critical aspects, particularly examining how sequential decision designs might influence users' evaluation of AI outputs.